\documentclass[prb,preprint,groupedaddress,showpacs]{revtex4}
\usepackage{graphicx} 
\begin{document} 
\bibliographystyle{prbrev}
 
\title{Vertically coupled double quantum rings at zero magnetic field}

\author{F. Malet}
\affiliation{Departament ECM, Facultat de F\'{\i}sica, Universitat de Barcelona. 
Diagonal 647, 08028 Barcelona, Spain}

\author{M. Barranco}
\affiliation{Departament ECM, Facultat de F\'{\i}sica, Universitat de Barcelona. Diagonal 647,
08028 Barcelona, Spain}

\author{E. Lipparini}
\altaffiliation{Permanent address:
Dipartimento di Fisica, Universit\`a di Trento, and INFN,
38050 Povo, Trento, Italy}
\affiliation{Departament ECM, Facultat de F\'{\i}sica, Universitat de Barcelona. Diagonal 647,
08028 Barcelona, Spain} 

\author{R. Mayol}
\affiliation{Departament ECM, Facultat de F\'{\i}sica, Universitat de Barcelona. Diagonal 647,
08028 Barcelona, Spain}

\author{M. Pi}
\affiliation{Departament ECM, Facultat de F\'{\i}sica, Universitat de Barcelona. Diagonal 647,
08028 Barcelona, Spain}

\author{J.I. Climente}
\affiliation{CNR-INFM S3 Center, Via Campi 213/A, 41100 Modena, Italy}

\author{J. Planelles}
\affiliation{Departament CC.EE., Universitat Jaume I. 12080
Castell\'o, Spain}

\date{\today}

\begin{abstract} 
Within local-spin-density functional theory, we have investigated the
`dissociation' of few-electron circular vertical semiconductor double
quantum ring artificial molecules at zero magnetic field as a function
of inter-ring distance. In a first step, the molecules are constituted
by two identical quantum rings. When the rings are quantum mechanically
strongly coupled, the electronic states are substantially delocalized,
and the addition energy spectra of the artificial molecule resemble
those of a single quantum ring in the few-electron limit. When the rings
are quantum mechanically weakly coupled, the electronic states in the
molecule are substantially localized in one ring or the other, although
the rings can be electrostatically coupled.
The effect of a slight mismatch introduced in the molecules from
nominally identical quantum wells, or from changes in the
inner radius of the constituent rings, induces localization by
offsetting the energy levels in the quantum rings. This plays a
crucial role in the appearance of the addition spectra as a function of
coupling strength particularly in the weak coupling limit. \\
\end{abstract}

%
%

\maketitle

\section{Introduction}

Semiconductor quantum dots (QD's) are widely regarded as artificial
atoms with properties similar to those of `natural' atoms.
One of the most appealling is the capability of forming molecules.
Systems composed of two QD's, quantum dot artificial molecules (QDM),
coupled either laterally or vertically, have been investigated
experimentally and theoretically at zero magnetic field ($B$), or
submitted to magnetic fields applied in different directions,
see e.g. Refs. \onlinecite{Pi01,Aus04,Pi05,Ron99,Par00}, and
references therein.

Semiconductor ring structures have also received considerable
attention in connection with the Aharonov-Bohm effect,\cite{Han01}
and the energy spectrum of nanoscopic self-assembled quantum rings
occupied by few electrons has been experimentally
analyzed.\cite{Lor00} Recently, high quality quantum rings (QR)
have been fabricated on a AlGaAs-GaAs heterostructure containing a
two-dimensional electron gas, by nano-litography with a scanning
force microscope, see Refs. \onlinecite{Fuh01} and 
\onlinecite{Ihn03}, and references therein. These studies have allowed
to extend previous studies to many-electron nanoscopic rings, and have
provided an experimental determination of the spin ground states
of the rings by Coulomb-blockade spectroscopy, as well as the clear
identification of a singlet-triplet transition, and the size of the
exchange interaction matrix element,\cite{Ihn05} properties that had
been also determined in the past for QD's.\cite{Tar96,Tar00}

Very recently, two different types of nanometer-sized QR complexes have
been realized. One such complex consists of two concentric QR's grown by
droplet epitaxy on an Al$_{0.3}$Ga$_{0.7}$As substrate.\cite{Kur05} The
other complex consists of stacked layers of InGaAs/GaAs QR's, whose
optical and structural properties have been characterized by
photoluminescence spectroscopy and by atomic force microscopy,
respectively.\cite{Sua04,Gra05}
Motivated by these recent experimental works, we have undertaken a
theoretical study, within local-spin-density functional theory (LSDFT),
 of the ground state (gs) properties of QR's complexes
at $B=0$. In this work, we present
the results we have obtained for the case of two GaAs vertical double
quantum rings, leaving aside for a separate study the case of
concentric double quantum rings, whose phenomenology is somewhat
different.\cite{ClimPRB} To some extent, our work parallels the one we
have carried
out in the past for double QD's, with the aim of understanding the
electronic properties of QRM's and the difference between 
vertical QDM and quantum ring molecule (QRM) structures.

This work is organized as follows. In Sec. II we present the
method we have used to describe QR's and vertically coupled QRM's.
The results we have obtained are discussed in Sec. III,
and a brief summary is presented in Sec. IV.

\section{LSDFT description of quantum rings and vertical quantum ring
molecules} 

We closely follow the method of Ref. \onlinecite{Pi01a}, where
the interested reader may find it described in some detail.
We recall that
within LSDFT, the ground state  of the system is
obtained by solving the Kohn-Sham equations.
The problem is simplified by the imposed axial symmetry 
around the $z$ axis, which allows one to write the
single particle (sp) wave functions as
$\phi_{nl\sigma}(r,z,\theta,\sigma)=
u_{nl\sigma}(r,z) e^{-\imath l \theta} \chi_{\sigma}$ with
$n =0, 1, 2, \ldots$, $l =0, \pm 1, \pm 2, \ldots$, being $-l$
the projection of the sp orbital angular momentum on the symmetry axis.

We have used effective atomic units 
$\hbar=e^2/\epsilon=m=$1, where 
$\epsilon$ is the dielectric constant,
and $m$  the electron effective mass. In units of the bare
electron  mass $m_e$ one has $ m = m^* m_e$.
In this system, the length unit is the effective
Bohr radius $a_0^* = a_0\epsilon/m^*$,
and the energy unit is the effective Hartree $H^* = H  m^*/\epsilon^2$.
In the numerical applications we have considered
GaAs, for which  we have taken $\epsilon$ = 12.4, and  $m^*$ = 0.067.
This yields $a^*_0 \sim$ 97.9 ${\rm \AA}$ and $H^*\sim$ 11.9 meV.

In cylindrical coordinates the KS equations read

\begin{eqnarray}
& & \left[-\frac{1}{2} \left( \frac{\partial^2}{\partial r^2}
+ \frac{1}{r} \frac{\partial}{\partial r} - \frac{l^2}{r^2} 
+ \frac{\partial^2}{\partial z^2} \right)
+ V_{cf}(r,z) \right.
\nonumber
\\
& &
\label{eq1}
\\
&+& \left. V^H + V^{xc} + W^{xc}\,\eta_{\sigma} \right]
u_{n l \sigma}(r,z) =
\varepsilon_{n l \sigma} u_{n l \sigma}(r,z) \,\, ,
\nonumber
\end{eqnarray}
where $\eta_{\sigma}$=$+1(-1)$ for $\sigma$=$\uparrow$$(\downarrow)$,
$V_{cf}(r,z)$ is the confining potential,
$V^H(r,z)$ is the direct Coulomb potential, and $V^{xc}={\partial
{\cal E}_{xc}(n,m)/\partial n}\vert_{gs}$ and
$W^{xc}={\partial
{\cal E}_{xc}(n,m)/\partial m}\vert_{gs}$
are the  variations of the exchange-correlation
energy density ${\cal E}_{xc}(n,m)$  in terms of the electron 
density $n(r,z)$ and of the local spin magnetization
$m(r,z)\equiv n^{\uparrow}(r,z)-n^{\downarrow}(r,z)$ taken at the 
gs.

As usual, 
${\cal E}_{xc}(n,m) \equiv {\cal E}_{x}(n,m) + {\cal E}_{c}(n,m)$
has been built from 3D homogeneous electron gas
calculations. This yields a well known,\cite{Lun83} simple analytical
expression for the exchange contribution ${\cal E}_{x}(n,m)$.
For the correlation contribution ${\cal E}_{c}(n,m)$ we have used
the parametrization proposed by
Perdew and Zunger.\cite{Per81}

For a double QR the confining potential $V_{cf}(r,z)$
has been taken parabolic in the $xy$ plane with a repulsive core around
the origin, plus
a symmetric double quantum well of width $w$ each, in the $z$ direction.
The potential in the $xy$ plane has circular symmetry, and in terms of
the cylindrical coordinate $r$ it is written as

\begin{equation}
V_{cf}(r) = 
V_0 \,\Theta(R_0-r) +
\frac{1}{2}\, m \,\omega_0^2\, (r - R_0)^2 \, \Theta (r-R_0) \;\; ,
\label{eq2}
\end{equation}
with $\Theta(x)=1$ if $x>0$ and zero otherwise. The convenience
of using a hard-wall confining potential to describe the effect of the inner 
core in QR's is endorsed by several works in the literature.\cite{Rin} 
We have taken $R_0=$ 5 nm, $w=$ 5 nm, $V_0=$ 350 meV, and $\omega_0$= 15
meV. The depth of the double quantum well is also $V_0$. This set of
parameters fairly represents the smallest rings synthesized in Ref.
\onlinecite{Lee04}, and
together with the distance $d$ between constituent quantum
wells, determine the confining potential. The distance $d$ is varied
to describe quantum ring molecules at different inter-ring distances.
For the  single `thick' QR  we will discuss as a reference system, we
have used the same confining potential in the $xy$ plane, together with a
single quantum well in the $z$ direction. For all structures, the sharp
potential wells have been slightly rounded off, as shown in Ref.
\onlinecite{Anc03}. Details about how the KS and Poisson eqs.
have been solved can be found in Ref. \onlinecite{Pi01a}.

\section{Results and discussion}
\subsection{Single quantum ring}
We have carried out calculations for a single thick QR
confined as indicated in the previous Section, and for a strictly
two-dimensional QR confined by the radial potential Eq. (\ref{eq2}),
as indicated, e.g., in Ref. \onlinecite{Emp01}.
These results will help us to discuss the appearence of the addition
spectra of the QRM.

\begin{figure}[t] 
\centerline{\includegraphics[width=07.5cm,angle=-90,clip]{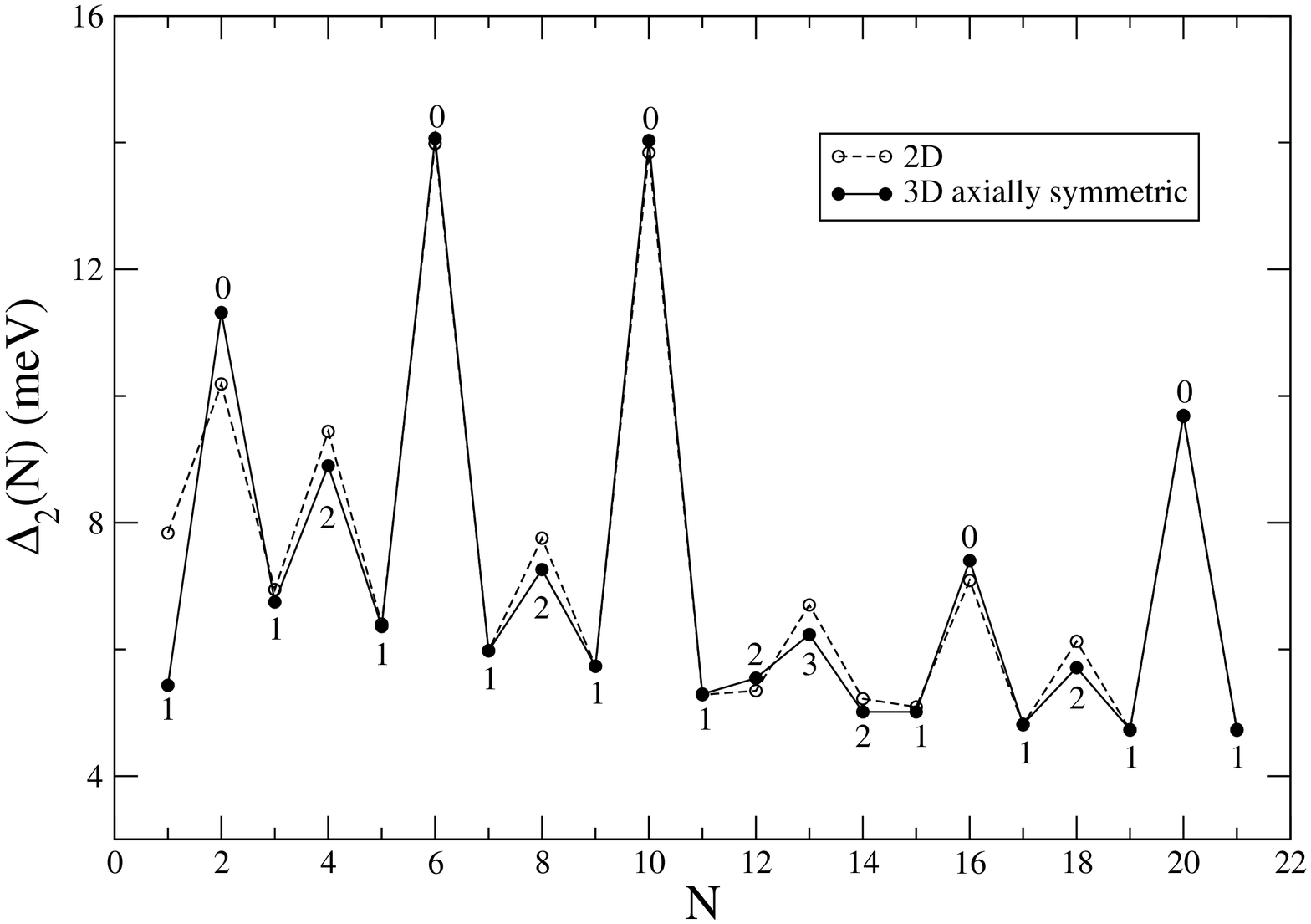} }
\caption{ 
Addition energies $\Delta_2(N)$ (meV) as a function of the number of
electrons $N$ for a thick QR (solid dots, solid lines) and a
strictly 2D QR (open dots, dashed lines). The value of $2S_z$ is
indicated.
} 
\label{fig1} 
\end{figure}

Fig. \ref{fig1} shows the addition energies $\Delta_2(N)$  

\begin{equation}
\Delta_2(N)= E(N+1)- 2 E(N)+ E(N-1)  \;\;\; ,
\label{eq3}
\end{equation}
where $E(N)$ is the total energy of the $N$ electron QR,
as a function of $N$.
It can be seen that the 2D and `thick' -i.e., axially symmetric 3D-
models sensibly yield the same
results for this observable, a well know result for QD's.\cite{Pi01} 
For the thick rings, the value of the calculated total spin third
component, $2S_z$, is also indicated in the figure.
We want to point out that in the $N=3$ case, the
2D model configuration is fully polarized $(2S_z=3)$. This is due to
the fact that the exchange-correlation energy is overestimated by
strictly 2D models.\cite{Ron99a} Fully polarized $N=3$ QR configurations
are not an artifact of the LSDFT. As a matter of fact, they have
been also found by exact diagonalization methods for some ring sizes and
confining potential choices.\cite{Zhu05}

The gs spin assignments we have found here coincide with those of Ref.
\onlinecite{Emp01}, although the height of the peaks in $\Delta_2(N)$
depends to a large extent on the confining potential. They are related
to the relative stability of the electronic shell closures in the ring,
which for $N>6$ are substantially different from these of QD's. In the
case of rings, they are mainly governed by the fourfold degeneracy of
the non-interacting sp levels with  $|l|\neq0$, and the
twofold degeneracy of the non-interacting sp levels with $|l|=0$.
This yields the marked shell
closures  at $N=2, 6, 10, 20$ and 28 with $S_z=0$, as well as the
$S_z=0$ gs found for $N=24$. The $2S_z=2$ ground states that
regularly appear between them indicate that Hund's rule is fulfilled 
by single QR's. 

\begin{figure}[t] 
\centerline{\includegraphics[width=07.5cm,angle=-90,clip]{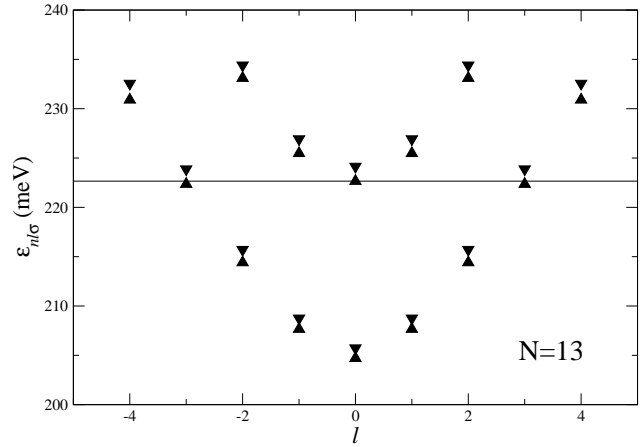} }
\caption{ 
Single particle  energy levels (meV) as a function of $l$ for a thick QR
with $N=13$.
Upward(downward) triangles denote $\uparrow$$(\downarrow$) spin states.
The thin horizontal line represents the Fermi energy.
} 
\label{fig2}
\end{figure}

The complex spin  structure around $N=13$ deserves some comments. It is due
to the occupancy of the second $(s\uparrow)$ state with $l=0$
-this spin structure is missing in other QR
calculations that employ a different confining potential.\cite{Lin01}
Figure \ref{fig2} displays the sp energies
$\varepsilon_{nl\sigma}$ for $N=13$, which are distributed
parabolic-like
as a function of $l$, each parabola corresponding to a different value
of the principal quantum number $n$. This figure explains
the filling sequence around $N=13$. For $N=12$, the second $(0\uparrow)$ 
state is empty, yielding $2S_z=2$; for $N=13$ the exchange interaction
favors the filling of this state yielding $2S_z=3$; for $N=14$, one of the 
$(\pm 3\downarrow)$
states is filled -they are degenerate-, yielding $2S_z=2$ (actually,
this many-electron configuration is nearly degenerate with the one in
which the $(0\downarrow)$ state is filled instead, which also yields
$2S_z=2$). For $N=16$, the
$(0\downarrow)$ and $(\pm3\downarrow)$ become populated, producing a
fairly strong shell closure.

\subsection{Homonuclear quantum ring molecules}

We consider first the case of a QRM formed by two
identical quantum rings. By analogy with natural molecules, we call
them homonuclear QRM. We have calculated their gs structure for $d=$ 2, 4
and 6 nm, and up to $N= 32$.  For a given electron number $N$, the
evolution of the gs (`phase') of a QRM as a function of $d$ may be thought
of as a dissociation
process.\cite{Pi01} Within LSDFT, each sp molecular
orbital has, as quantum labels, the third component of the spin and of
the orbital angular
momentum, the parity, and the value of reflection symmetry about the
$z=0$ plane. Symmetric states $|S\rangle$ are called bonding states, and
antisymmetric states $|AS\rangle$ are called antibonding states.

The energy splitting between bonding and antibonding sets of sp states,
$\Delta_{SAS}$, can be properly estimated\cite{Pi01} from the energy
difference of the antisymmetric and symmetric  states of a single
electron QRM,  $\Delta_{SAS}\sim E(^2$$\Sigma^-_u)-$$E(^2$$\Sigma^+_g)$
-see below for the notation-,
and varies from 24.9 meV at $d=2$ nm (strong coupling), to
1.49 meV at $d=6$ nm (weak coupling). 
In this range of inter-ring distances, $\Delta_{SAS}$ can be fitted as
$\Delta_{SAS}=\Delta_0\,e^{-d/d_0}$, with $\Delta_0=82$ meV and $d_0=1.68$ nm.
The relative value of the two energies $\hbar \omega_0$
and $\Delta_{SAS}$  crucially determines the structure of the molecular
phases along the dissociation path.

\begin{figure}[t] 
\centerline{\includegraphics[width=10.5cm,clip]{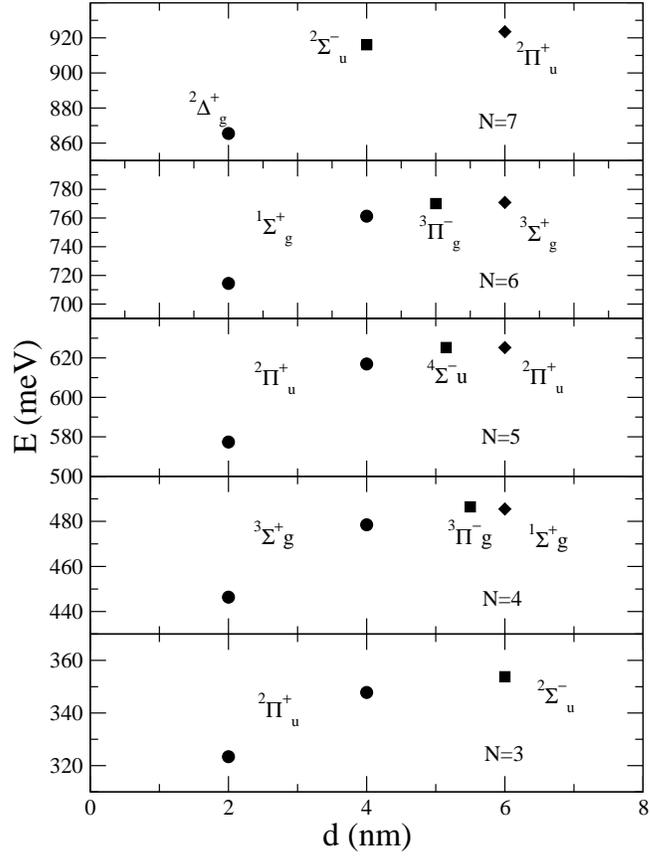} }
\caption{ 
Energy (meV) and gs molecular phases of the homonuclear QRM
as functions of the inter-ring distance $d$ for $N=3-7$.
For each QRM, different phases are represented by different symbols.
} 
\label{fig3}
\end{figure}

Figure \ref{fig3} shows the evolution with $d$ of the gs energy  and
molecular phase of a QRM made of $N=3-7$ electrons.
Each configuration  is labeled using an adapted version of the
ordinary spectroscopy notation,\cite{Ron99} namely
$^{2S+1}L^{\pm}_{g,u}$, where $S$ is the total $|S_z|$, and $L$ is the
total $|L_z|$. The superscript $+(-)$ refers to even (odd) states under
reflection with respect to the $z=0$ plane, and the subscript $g(u)$
refers to positive(negative) parity states. To label the molecular sp
states we have used the standard convention of molecular physics, using
$\sigma, \pi, \delta, \ldots$ if $l=0, \pm 1, \pm 2, \ldots$. Upper case Greek
letters are used for the total $|L_z|$.
Fig. \ref{fig3} shows that the energy of the molecular
phase increases with $d$. This is due to the increase  of the
energy of the sp bonding states as $d$ increases,\cite{Pi01a} that
dominates over the decrease in Coulomb energy. At
larger inter-ring distances, the constituent QR are so apart that
eventually
the decrease of Coulomb energy dominates and the tendency is reversed.
The phase sequences are the same as for double quantum dots,\cite{Pi01}
although the transition inter-ring distances, which obviously depend on
the kind
and strength of the confining potential, are different. As for double
quantum dots, we have found that the first phase transition of a few-electron 
QRM is always due to the replacement of an occupied bonding sp
state by an empty antibonding one. 

\begin{figure}[t] 
\centerline{\includegraphics[width=10.5cm,clip]{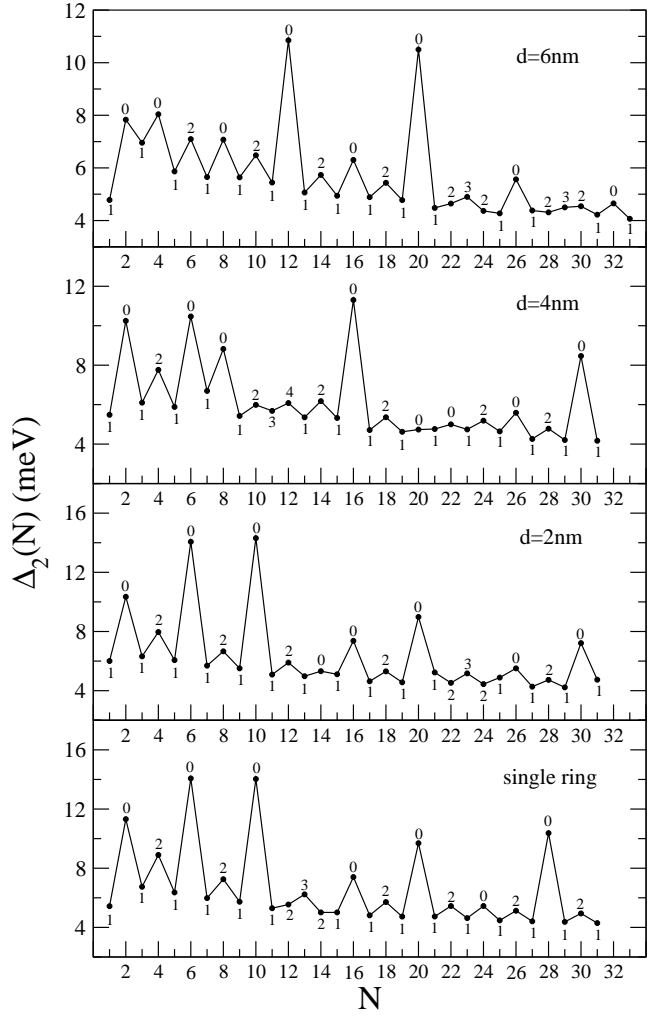} }
\caption{ 
$\Delta_2(N)$ for homonuclear QRM with inter-ring
distances $d=2, 4$, and 6 nm. 
The addition energies have been offset for clarity.
Also shown is the reference spectrum for a single QR.
The value of $2S_z$ is indicated.
} 
\label{fig4}
\end{figure}

Figure \ref{fig4} shows the addition spectra for homonuclear QRM up to
$N=31$ for the three selected inter-ring distances. Also shown is the
reference spectrum of a single QR. The spectra have been offset for
clarity. For small $d$ ($\Delta_{SAS} \gg \hbar \omega_0$) the spectrum
of the QRM is rather similar to a single QR, especially for few-electron
systems, with minor changes arising
in the $N \sim 12$ and $\sim 24$ regions that will be commented below.
It is clear that for $d=2$ nm the two QR's are electrostatically and
quantum-mechanically coupled and behave as a single system. 
At intermediate distances the spectrum pattern becomes more complex, but at
larger distances (e.g. $d=6$ nm), when the QRM molecule is about to
dissociate, the physical picture that emerges is rather simple 
and can be interpreted using intuitive, yet approximate arguments:
At large distances ($\Delta_{SAS} \ll \hbar \omega_0$),
 the QR's are coupled only electrostatically, and
most $|S\rangle$ and $|AS\rangle$ states are quasidegenerate.
Electron localization in each constituent QR can be
achieved combining these states as $(|S\rangle \pm|AS\rangle)/\sqrt{2}$
and as a consequence, the strong $S_z=0$ peaks found at
$N=12$ and 20 are readily interpreted from the peaks appearing in the
single QR spectrum at $N=6$ and 10; the process can be viewed as the
symmetric dissociation of the original QRM leading to very robust
closed shell single QR configurations. This is also the origin of
the QRM $S_z=0$ peaks at $N=2$ and 4. In the former case, the QRM
configuration
corresponds to one single electron being hosted in each constituent QR
coupled into a singlet state, and in the latter case, the QRM
configuration is viewed as two QR's, each one occupied by two electrons
filling the $1s$ shell.

\begin{figure}[t] 
\centerline{\includegraphics[width=10.5cm,clip]{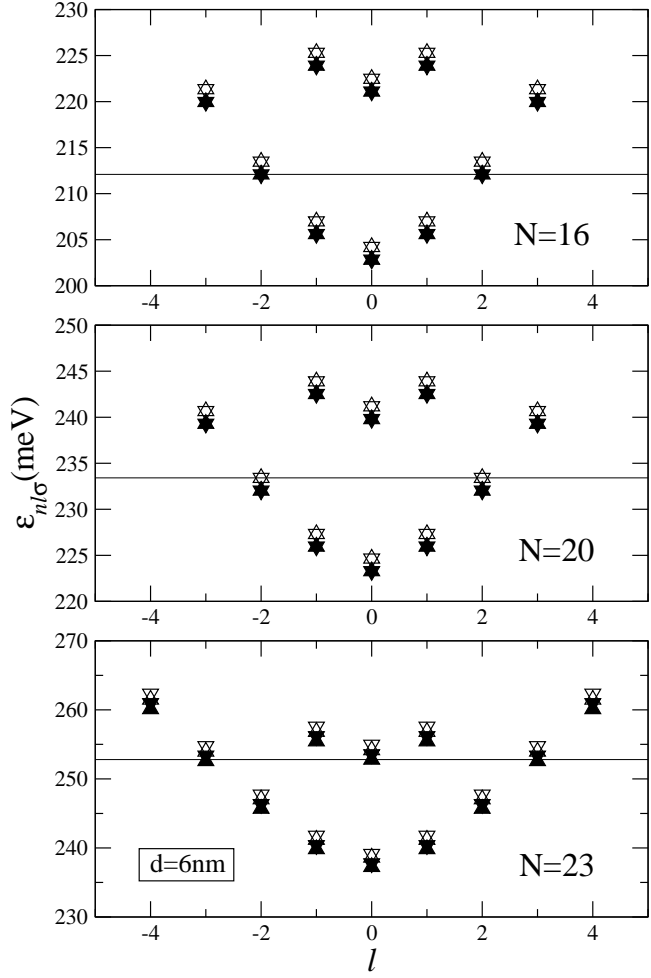} }
\caption{ 
Single particle  energy levels (meV) as a function of $l$ for 
an homonuclear QRM with $d=6$ nm  and
$N=16$ (top panel), $N=20$ (middle panel), and $N=23$ (bottom
panel) .
Upward(downward) triangles denote $\uparrow$($\downarrow$) spin states.
Open(solid) triangles correspond to antibonding(bonding) states.
The thin horizontal line represents the Fermi energy.
} 
\label{fig5}
\end{figure}

At $d=6$ nm, other dissociations display a more complicated pattern,
such as $16 \rightarrow 8 +8$,  or $8 \rightarrow 4 +4$, whose final
products are QR's that fulfill Hund's rule whereas the actual
QRM has $S_z=0$. These could
be interpreted as rather entangled QRM, `harder' to dissociate, for
which a $d=6$ nm inter-ring distance is not large enough 
to allow for electron localization. The fact is
that not only quasidegeneracy of occupied $|S\rangle$ and $|AS\rangle$
states at given $d$ plays a role
in this intuitive analysis, but also whether their number is equal or
not, so that they may be
eventually combined to favor localization. An example of these two
different situations is illustrated in Fig. \ref{fig5}, where we show
the sp states of the $N=16$, 20 and 23 QRM at $d=6$ nm. In the case of
$N=16$ and 23, the filled bonding states near the Fermi level have not
filled antibonding partner and are delocalized in
the whole volume of the QRM, contributing to the molecular bonding at
that
distance, whereas all other bonding states can be localized combining
them with their antibonding partner: as in natural molecules, some
orbitals contribute to the molecular bonding, whereas some others do
not.

\subsection{Heteronuclear quantum ring molecules}

For vertically coupled lithographic double quantum dots, it has been found
unavoidable that a slight mismatch is unintentionally introduced in 
the course of their fabrication from materials with nominally identical
constituents quantum wells,\cite{Pi01} which is responsible for electron
localization as the interdot coupling becomes weaker. This offsets the
energy levels in the quantum dots by a certain amount that was there
estimated to be up to 2 meV, and this plays a crucial role in the
appearance of the addition energy spectra as a function of the coupling
strength particularly in the weak coupling limit. A similar picture 
is also found in coupled self-assembled quantum dots, where strain 
propagation between adjacent layers of dots often leads to top QD's of
increased size.\cite{Le96}

Likely, the same fabrication limitations will appear in the case of
vertically coupled double quantum rings. Anticipating to this situation,
we have carried out a series of QRM calculations in which the double
quantum wells have the same width $w$ but slightly different depths,
namely $V_0\pm\delta$, with $\delta \ll V_0$. It can be easily checked that
in the weak coupling limit ($2\delta \gg \Delta_{SAS}$), $2\delta$ is
approximately the energy splitting between the bonding and antibonding
sp states, which would be almost degenerate if $\delta=0$. For
this reason we call the mismatch (offset) the quantity $2\delta$.

\begin{figure}[t] 
\centerline{\includegraphics[width=10.5cm,clip]{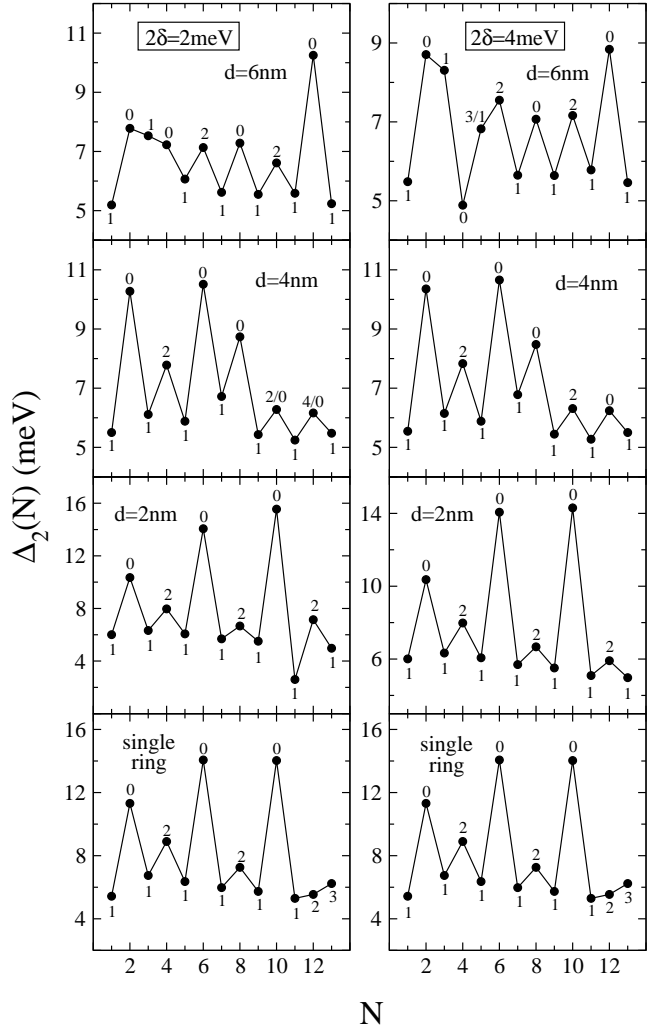} }
\caption{ 
$\Delta_2(N)$ for heteronuclear QRM with inter-ring
distances $d=2, 4$, and 6 nm and mismatch $2\delta=2$ meV
(left panels) and  $2\delta=4$ meV (right panels).
The addition energies have been offset for clarity.
Also shown is the reference spectrum for a single QR.
The value of $2S_z$ is indicated.
Note that in some cases two different values of $S_z$ have been
asigned to the same peak. This means that the corresponding
configurations are nearly degenerated.
} 
\label{fig6}
\end{figure}

We have considered two possible values of the mismatch, namely
$2\delta=2$ and 4 meV, and have obtained the corresponding addition
spectra for up to $N=13$ electrons -according to our previous
experience with double quantum dots,\cite{Pi01}  we
expect that the larger differences will arise in few-electron QRM.
The results are displayed in Fig. \ref{fig6}. It can be seen that in the
strong coupling limit, the effect of the mismatch on the addition
energies is negligible, as expected.\cite{Pi01} The electrons are
completely delocalized in the whole volume of the QRM, and the
introduced mismatch is unable to localize them in either of the
constituents QR's. 

For the few-electron QRM, which is the more interesting physical
situation, we have shown before that the fingerprint of homonuclear
character is the appearance,
in the weak coupling limit, of the peaks in the addition spectrum
corresponding to $N=2$ and 4, as well as the spin assignment $S_z=0$.
It can be seen from Fig. \ref{fig6} that in the intermediate regime
($d=4$ nm) the $N=4$ peak still corresponds to a $2S_z=2$ configuration,
but at larger inter-ring distances, it eventually
disappears, yielding an addition spectrum that clearly manifests the
heteronuclear character of the QRM and constitutes a clean fingerprint
of these kind of configurations.

\begin{figure}[t] 
\centerline{\includegraphics[width=10.5cm,clip]{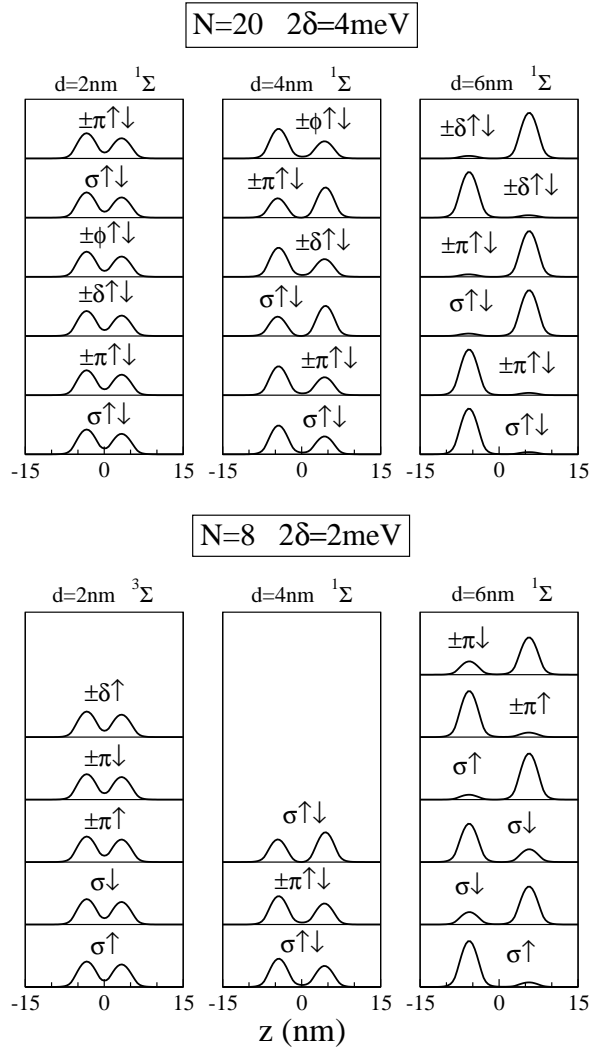} }
\caption{ 
Calculated probability distributions ${\cal P}(z)$ (arbitrary units) as
a function of $z$ for heteronuclear QRM's with $N=20$, $2\delta=4$ meV
(top panels), and  $N=8$, $2\delta=2$ meV (bottom panels). The
corresponding molecular configuration is also indicated.}
\label{fig7}
\end{figure}

It is useful to display the dissociation of the QRM representing the $d$
evolution of the sp molecular wave functions, introducing the
$z$-probability distribution function\cite{Pi01}

\begin{equation}
{\cal P}(z) = 2 \pi \int dr \,r \, [u(r,z)]^2  \;\;\; .
\label{eq4}
\end{equation}
Examples of these probability functions can be seen in Fig.
\ref{fig7}, where we show ${\cal P}(z)$ for $N=20$, $2\delta=4$ meV,
and $N=8$, $2\delta=2$ meV (deeper well always in the $z<0$ region),
each for the chosen three $d$ values. In
each panel the probability functions
are plotted ordered from bottom to top according to the increasing sp
energies. For $N=20$ the final configurations are the closed shell
$N=10$, $2S_z=0$ QR's, whereas for $N=8$ the  $N=4$,
$2S_z=2$ Hund's rule QR configurations emerge.

Finally, we discuss the case of two QR's of different radii vertically 
coupled  to build an axially symmetric QRM,
and  study the effect this assymetry  has on the addition spectrum
(we have discarded a possible disalignment of the QR symmetry
axes, as addressing this situation would require a much more
demanding full 3D calculation\cite{Pi04}).
To this end, we have taken for one ring  $R_0=6$ nm, while
for the other one we have kept the same value as before, $R_0=5$ nm 
($\delta$ is set to zero this case). 

\begin{figure}[t] 
\centerline{\includegraphics[width=10.5cm,clip]{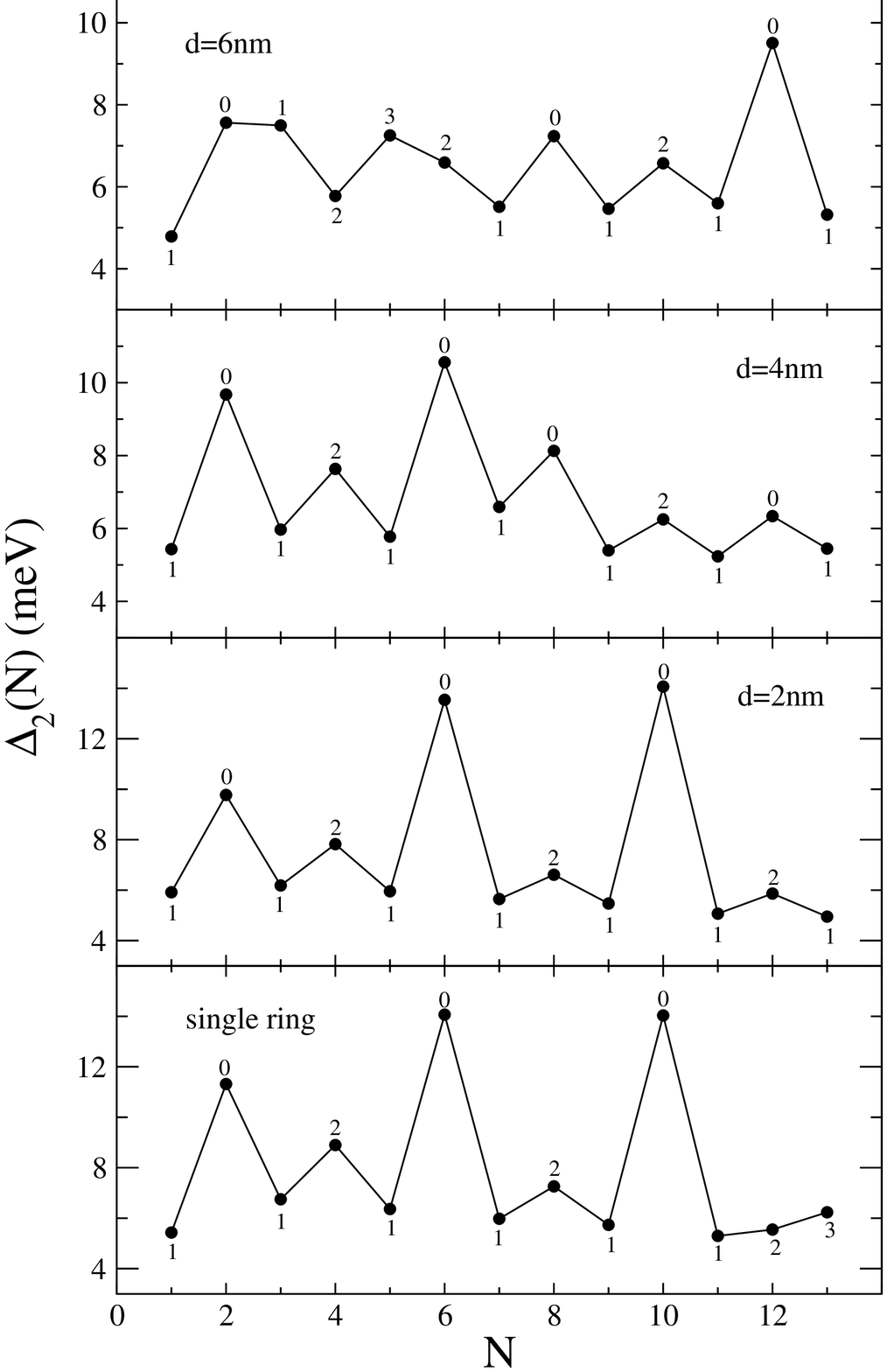} }
\caption{ 
$\Delta_2(N)$ for heteronuclear QRM made of QR with different core
radii $R_0=5$ and $R_0=6$ nm, hosting up to $N=14$
electrons and inter-ring
distances $d=2, 4$, and 6 nm.
The addition energies have been offset for clarity.
Also shown is the reference spectrum for a single QR.
The value of $2S_z$ is indicated.
} 
\label{fig8}
\end{figure}

We show in Fig. \ref{fig8} the addition spectra for up to $N=14$
electrons and $d=2$, 4, and 6 nm. It can be seen that in the strong and
intermediate coupling cases they are fairly similar to the previous
heteronuclear case -and to the homonuclear case as well-, indicating a
fairly robust structure of the QRM in these limits.
As before, the heteronuclear character clearly shows up in the weak
coupling limit, with a peak structure and $S_z$ assignments remarkably
similar to those discussed in the previous situation with $\delta \neq
0$.

\section{Summary and outlook}

We have discussed the appearance of the addition energy spectra of
homonuclear and heteronuclear
quantum ring molecules at zero magnetic field. In particular, we
have addressed the addition energy spectrum of QRM from the weak to the
strong coupling limits. Fingerprints of homo- and heteronuclear molecular
character
have been pointed out in the weak coupling limit. As it happened in the
study of vertically coupled double quantum dots,\cite{Pi01} we believe
this may be helpful in the analysis of future experiments on vertical QRM's.

The present study can be naturally extended to the case of QRM's submitted 
to magnetic fields of arbitrary direction. A rich interplay between molecular
phases having different isospin is expected to appear as a function of
$B$,\cite{Anc03,Aus04} which might have an observable influence on the 
Aharonov-Bohm effect and on the far-infrared spectroscopy of nanoscopic
QRM's.

\section*{ACKNOWLEDGMENTS} This work has been performed 
under grants FIS2005-01414 from 
DGI (Spain), 2005SGR00343 from Generalitat de Catalunya, and
CTQ2004-02315/BQU, UJI-Bancaixa Contract No. P1-B2002-01 (Spain).
E. L. has been suported by DGU (Spain), grant SAB2004-0091, and by
CESCA-CEPBA, Barcelona, in the initial stages of this work
through the program HPC-Europa Transnational Access.
J.I.C. has been supported by the EU under the TMR network `Exciting'.

\pagebreak

\pagebreak

\pagebreak

\pagebreak

\pagebreak

\pagebreak

\pagebreak

\pagebreak

\begin{thebibliography}{99}

\bibitem{Pi01} M. Pi, A. Emperador, M. Barranco, F. Garcias, K. Muraki,
S. Tarucha, and D.G. Austing, Phys. Rev. Lett. {\bf 87}, 066801 (2001).

\bibitem{Aus04} D.G. Austing, S. Tarucha, H. Tamura, K. Muraki, F.
Ancilotto, M. Barranco, A. Emperador, R. Mayol, and M. Pi,
Phys. Rev. B {\bf 70}, 045324 (2004).

\bibitem{Pi05} M. Pi, D.G. Austing, R. Mayol, K. Muraki,
S. Sasaki, H. Tamura, and S. Tarucha,  in {\it Trends in Quantum
Dots Research}, P. A. Ling, editor p. 1 (Nova Science Pu. 2005).

\bibitem{Ron99} M. Rontani, F. Rossi, F. Manghi, and E. Molinari, Solid
State Commun. {\bf 112}, 151 (1999)

\bibitem{Par00} B. Partoens and F.M. Peeters, Phys. Rev. Lett. 
{\bf 84}, 4433 (2000); Europhys. Lett. {\bf 56}, 86 (2001).

\bibitem{Han01} A.E. Hansen, A. Kristensen, S. Pedersen, C.B. Sorensen,
and P.E. Lindelof, Phys. Rev. B {\bf 64}, 45327 (2001).

\bibitem{Lor00} A. Lorke, R.J. Luyken, A.O. Govorov, J.P. Kotthaus, J.M. Garcia,
and P.M. Petroff, Phys. Rev. Lett. {\bf 84}, 2223 (2000).

\bibitem{Fuh01} A. Fuhrer, S. L\"uscher, T. Ihn, T. Heinzel,
K. Ensslin, W. Wegscheider, and M. Bichler, Nature {\bf 413}, 822
(2001).

\bibitem{Ihn03} T. Ihn, A. Fuhrer, T. Heinzel, K. Ensslin, W.
Wegscheider, and M. Bichler, Physica E {\bf 16}, 83 (2003).

\bibitem{Ihn05} T. Ihn, A. Fuhrer, K. Ensslin, W.
Wegscheider, and M. Bichler, Physica E {\bf 26}, 225 (2005).

\bibitem{Tar96} S. Tarucha, D.G. Austing, T. Honda,  R.J. van
der Hage, and L.P. Kouwenhoven, Phys. Rev. Lett. {\bf 77}, 3613 (1996).

\bibitem{Tar00} S. Tarucha, D.G. Austing, Y. Tokura, W.G. van der Wiel,
and L.P. Kouwenhoven, Phys. Rev. Lett. {\bf 84}, 2485 (2000).

\bibitem{Kur05} T. Kuroda, T. Mano, T. Ochiai, S. Sanguinetti, K.
Sakoda, G. Kido, and N. Koguchi, Phys. Rev. B {\bf 72}, 205301 (2005).

\bibitem{Gra05} D. Granados, J.M. Garc\'{\i}a, T. Ben, and S.I. Molina,
Appl. Phys. Lett. {\bf 86}, 071918 (2005).

\bibitem{Sua04} F. Su\'arez, D. Granados, M.L. Dotor, and J.M.
Garc\'{\i}a, Nanotechnology {\bf 15}, S126 (2004).

\bibitem{ClimPRB} J.I. Climente, J. Planelles, M. Barranco, F.
Malet, and M. Pi, unpublished (2006).

\bibitem{Pi01a} M. Pi, A. Emperador, M. Barranco, and F. Garcias,
Phys. Rev. B {\bf 63}, 115316 (2001).

\bibitem{Lun83} S. Lundqvist,
{\em Theory of the Inhomogenous
Electron Gas}, edited by S. Lundqvist and N. H. March
(Plenum, New York, 1983) p. 149.
 
\bibitem{Per81} J. P. Perdew and A. Zunger,
Phys. Rev. B {\bf 23}, 5048 (1981).

\bibitem{Anc03} F. Ancilotto, D.G. Austing, M. Barranco, R. Mayol, K.
Muraki, M. Pi, S. Sasaki, and S. Tarucha, Phys. Rev. B {\bf 67}, 205311 (2003).

\bibitem{Rin} S.S. Li, and J.B. Xia, J. Appl. Phys. {\bf 89}, 3434 (2001);
A. Puente, and Ll. Serra, Phys. Rev. B {\bf 63}, 125334 (2001);
J.I. Climente, J. Planelles, and F. Rajadell, J. Phys.: Condens.  
Matter {\bf 17}, 1573 (2005).

\bibitem{Lee04} B.C. Lee and C.P. Lee, Nanotechnology {\bf 15},
848 (2004).

\bibitem{Emp01} A. Emperador, M. Pi, M. Barranco, and E. Lipparini,
Phys. Rev. B {\bf 64}, 155304 (2001).

\bibitem{Ron99a} M. Rontani, F. Rossi, F. Manghi, and E. Molinari, Phys.
Rev. B {\bf 59}, 10165 (1999).

\bibitem{Zhu05} J-L. Zhu, S. Hu, Z. Dai, and X. Hu,
Phys. Rev. B {\bf 72}, 075411 (2005).

\bibitem{Lin01} J.C. Lin and G.Y. Guo,
Phys. Rev. B {\bf 65}, 035304 (2001).

\bibitem{Le96} N.N. Ledentsov, V.A. Shchukin, M. Grundmann, N. Kirstaedter, 
J. B\"ohrer, O. Schmidt, D. Bimberg,V.M. Ustinov, A. Yu. Egorov, A.E.
Zhukov, P.S. Kop'ev, S.V. Zaitsev, N.Yu. Gordeev, Zh.I. Alferov, A.I. Borovkov, 
A.O. Kosogov, S.S. Ruvimov, P. Werner, U. G\"osele, and J. Heydenreich,
 Phys. Rev. B {\bf 54}, 8743 (1996). 

\bibitem{Pi04} M. Pi, F. Ancilotto, E. Lipparini, and R. Mayol,
Physica E {\bf 24}, 297 (2004).


\end{thebibliography}
\end{document}